\documentclass[12pt]{article}
\usepackage{amsmath}
\usepackage{amssymb}
\usepackage{amsfonts}
\usepackage{latexsym}
\usepackage{color}
\usepackage{graphicx}

\catcode `\@=11 \@addtoreset{equation}{section}

\catcode `\@=12

\newcommand{\be}{\begin{equation}}
\newcommand{\en}{\end{equation}}
\newcommand{\bea}{\begin{eqnarray}}
\newcommand{\ena}{\end{eqnarray}}
\newcommand{\beano}{\begin{eqnarray*}}
\newcommand{\enano}{\end{eqnarray*}}
\newcommand{\bee}{\begin{enumerate}}
\newcommand{\ene}{\end{enumerate}}

\newcommand{\Hil}{{\cal H}}

\newcommand{\Id}{1\!\!1}

\newcommand{\Lc}{{\cal L}}
\newcommand{\Sc}{{\cal S}}

\newcommand{\A}{{\cal A}}
\newcommand{\1}{1 \!\! 1}

\newcommand{\ltwo}{{\Lc^2(\mathbb{R})}}

\renewcommand{\l}{\langle}
\renewcommand{\r}{\rangle}
\newcommand{\pin}[2]{\l#1 , #2\r}

\begin{document}
\thispagestyle{empty}

\vspace*{1cm}

\begin{center}
{\Large \bf A new family of ladder operators for macroscopic systems, with applications}   \vspace{2cm}\\

{\large F. Bagarello}
\vspace{3mm}\\[0pt]
Dipartimento di Ingegneria,\\[0pt]
Universit\`{a} di Palermo, I - 90128 Palermo,  and\\
\, INFN, Sezione di Catania, Italy.\\[0pt]
E-mail: fabio.bagarello@unipa.it\\[0pt]
\end{center}

\vspace*{0.5cm}

\begin{abstract}
	In a series of recent scientific contributions the role of bosonic and fermionic ladder operators in a macroscopic realm has been investigated. Creation, annihilation and number operators have been used in very different contexts, all sharing the same common main feature, i.e. the relevance of {\em discrete changes} in the description of the system. 
	The main problem when using this approach is that computations are easy for Hamiltonians which are quadratic in the ladder operators, but become very complicated, both at the analytical and at the numerical level, when the Hamiltonian is not quadratic. In this paper we propose a possible alternative approach, again based on some sort of ladder operators, but for which an analytic solution can often be deduced without particular difficulties. We describe our proposal with few applications, mostly related to different versions of a predator-prey model, and to love affairs (from a decision-making point of view).

\end{abstract}

\vspace{2cm}


\vfill

\newpage

\section{Introduction}\label{sect1}

In the past few decades, in a series of papers and books, quantum tools and quantum ideas have been used in connection with systems which are not apparently connected with quantum mechanics at all. For instance, biology, economy, psychology, are only few of the realms of the research which have been considered, and still are considered, using this quantum-like approach. We refer to \cite{qal}-\cite{buse} for some books and few recent papers, where several other references can be found.

One of the main critical aspects of this approach is possibly {\em why} one should use, for instance, operators and vectors on some Hilbert space instead of ordinary functions, differential equations, and a deterministic (or a stochastic) approach,  which are much more diffused

\noindent in the scientific literature. Our personal reply to this objection is that, in some particular situations, operators of some special kind work very well in the analysis of the system, and that the model it is better constructed in terms of, say, ladder operators rather than of continuously changing functions. This is mainly because there are several situations in which the relevant observables (units of cash, number of shares, members of a populations, people moving from a place to another,...) which change discontinuously, and using continuous functions looks only a matter of convenience.

Also, there are quantum ideas (like, to cite two, the uncertainty principle and the general noncommutativity of the observables) which are surely relevant also in macroscopic situations, and not only in the microscopic world. For instance, Decision Making is one such case, in particular when it is needed to take decisions under uncertainty, \cite{yuka}, or in order to explain order effects in some poll, \cite{paulina,buse}.



However, as it is clear from the existing literature on this approach, it is often much more complicated, at a purely technical level, to deal with operators rather than with functions. This has forced people often to oversimplify the model proposed to describe the system $\Sc$ under analysis, and in particular the Hamiltonian operator needed in the deduction of the dynamical behaviour of $\Sc$, \cite{bagbook,bagbook2,FFF}. In fact, quite often, only quadratic Hamiltonians have been considered in the literature. The rationale for this choice is that, in this way, the equations of motion for the ladder operators, deduced using the Heisenberg recipe, became linear, and can be solved exactly. However, in many situations, it is easy to imagine that the realistic Hamiltonian should not be quadratic. For instance, when three agents interact, we expect that the operators of each agent should somehow appear together. This gives rise to a cubic term in the Hamiltonian, and to a related non linear differential equation for the ladder operators, whose solution is not easy, at least at an analytic level. Cubic contributions can be easily taken into account if the ladder operators are fermionic, \cite{bagbook,bagbook2,FFF}, at the (modest) price to work in a 8-dimensional Hilbert space: here $8=2^3$ is the dimension of the {\em smaller} Hilbert space where 3 modes of fermionic operators can be represented. In other words, every raising and lowering operator of a 3-agents system can be represented as a $8\times 8$ matrix, acting on vectors with 8 complex-valued components. The situation becomes more and more complicated when the number of agents increases. This is unavoidable if we are interested, for instance, to spatial models where different agents (the {\em actors} in $\Sc$) may occupy different cells of a given lattice describing the region where the system {\em lives}. In this case, in presence of $L$ different agents diffused in $M$ cells, we have to deal with $2^{M\times L}\times2^{M\times L}$ matrices. This means that the numerical difficulties become harder and harder since the number of coupled differential equations to be solved simultaneously increases very fast, and makes the model impossible to be dealt without extremely powerful computers. The situation is even more complicated when fermions are replaced by bosons, since in this case we need to deal, already for purely quadratic Hamiltonians, with infinite-dimensional Hilbert spaces. So, in absence of any {\em simplifying assumption}, there is no easy way to deal with Hamiltonians written in terms of bosonic operators which are not quadratic. This is essentially because there is no finite dimensional representation of the bosonic ladder operators, so that, in principle, we should solve a set of infinite differential equations. On the other hand, if the Hamiltonian $H$ is quadratic, the solution can be easily computed (at a formal level), and the difficulty is that of computing mean values, and to store quantities (a huge number of them) on our computers.  Of course, dealing only with quadratic Hamiltonians can really appear unsatisfying, when dealing with concrete systems. For this reason, in \cite{therule,HRO2} the so-called $(H,\rho)$-induced dynamics was introduced. This is a way to modify, again and again, the time evolution of  $\Sc$ by letting the system evolve by means of a (quadratic) Hamiltonian $H$ in a certain time interval, $[0,\tau[$. At $t=\tau$ a {\em check} is performed on $\Sc$, and the system is modified, in part, as a result of this check. This is what we call {\em the rule} $\rho$. Then $\Sc$ evolves again according to $H$ for another time interval of length $\tau$, when a second check on $\Sc$ is performed. And so on. This approach has proved to have many interesting applications, and to be interesting both at a mathematical level and for its possible use in concrete systems. However, it should be stressed that the numerical difficulty, thought being somehow under control, remains still quite high, mainly because we have to take into account the effect of the rule $\rho$ after each time interval of length $\tau$, which might be not a simple task.

Summarizing, we conclude that the use of bosonic and fermionic operators can help us to construct interesting Hamiltonian operators describing efficiently $\Sc$, quadratic or not, and we can derive the equations of motion out of them. However, the solution of these equations is not easy, in general, and the use of some other techniques or ladder operators can be relevant to enlarge the range of applicability of our general idea, which is to deduce equations of motion not from the Netwon's but from the Heisenberg's equation of motion. This is exactly what we will explore in this paper: we will propose a different approach, again related to ladder operators, which allow a fully analytical solution of the dynamical problem for a given system $\Sc$ even in presence of non quadratic Hamiltonians, at least for the quite large class of Hamiltonians we will consider in the rest of this paper.

The paper is organized as follows: in the next section we briefly review the role of fermionic and bosonic ladder operators, and we mention few applications which have been considered along the years. In Section \ref{sect3} we introduce a different class of ladder operators, listing some useful rules and some of their properties. In Section \ref{sect4}  we use these new operators to construct a first toy model which helps us in understanding which are the differences between these latter and the previous (fermionic or bosonic) ladder operators. In Section \ref{sect5} we propose a sort of cubic predator-prey model, with three different agents, which is relevant to show that more complicated Hamiltonians, not necessarily quadratic in the ladder operators, can be easily considered within our new scheme. Then, in Section \ref{sect6}, we consider a much more complicated model, which is still cubic, but with only two agents. The difference of this model with the one considered in Section \ref{sect4} is that, now, we  introduce a sort of control which {\em activates} or not the interaction between the agents. Section \ref{sect7} contains our conclusions, and some comments on how to adapt the models considered all along the paper to a completely different problem, and in particular to  {\em Decision Making} and, more in details, to a love affair between Alice and Bob, on the same lines already considered in \cite{bagoli,bag5}.

\section{Standard ladder operators}\label{sect2}

In this section we briefly review few facts on fermionic and bosonic ladder operators, and we mention few systems which have been described in terms of them.

\subsection{Fermionic operators}

Given a set of operators  $\{b_l,\,b_l^\dagger, \ell=1,2,\ldots,L\}$ acting on a certain \index{Hilbert space}Hilbert space $\Hil_F$, we say
that they satisfy the {\em canonical anticommutation relations} (CAR) if the conditions \be
\{b_l,b_n^\dagger\}=\delta_{l, n}\Id,\hspace{8mm} \{b_l,b_n\}=\{b_l^\dagger,b_n^\dagger\}=0 \label{a3} \en hold true for all
$l,n=1,2,\ldots,L$. Here,  $\{x,y\}:=xy+yx$ is the {\em anticommutator} of $x$ and $y$ and $\Id$ is the identity
operator on $\Hil_F$. These operators, which are considered in many textbooks on quantum mechanics, e.g. in \cite{mer,mess}, are
those which are used to describe $L$ different {\em modes} of fermions. From these operators we can
construct the Hermitian operators $\hat n_l=b_l^\dagger b_l$ and $\hat N=\sum_{l=1}^L \hat n_l$. In particular, $\hat n_\ell$ is the
{\em number operator} for the $\ell$--th mode, while $\hat N$ is the \emph{global number operator}. It is interesting to notice that the operators in (\ref{a3}) satisfy a very important feature: if we try to square them (or to rise them to higher powers), we
simply get zero: for instance,  (\ref{a3}) implies that $b_{l}^2=0$. This is of course related to the fact that
fermions satisfy the Pauli exclusion principle, \cite{mer,mess}.

The Hilbert space of our system is constructed introducing first the \emph{vacuum} of the theory, that is a vector
$\Phi_{\bf 0}$ which is annihilated by all the operators $b_l$: $b_l\Phi_{\bf 0}=0$ for all $l=1,2,\ldots,L$. Then we act on $\Phi_{\bf 0}$
with the  operators $(b_l^\dagger)^{n_l}$: \be \Phi_{n_1,n_2,\ldots,n_L}:=(b_1^\dagger)^{n_1}(b_2^\dagger)^{n_2}\cdots
(b_L^\dagger)^{n_L}\Phi_{\bf 0}, \label{a4} \en $n_l=0,1$, for all $l$. Of course, we do not consider higher powers of the $b_j^\dagger$'s since
these powers would simply destroy the vector.  These vectors form an orthonormal (o.n.) set which spans 
$\Hil_F$, and they are eigenstates of both $\hat n_l$ and $\hat N$:
$$\hat
n_l\Phi_{n_1,n_2,\ldots,n_L}=n_l\Phi_{n_1,n_2,\ldots,n_L}$$ and
$$\hat N\Phi_{n_1,n_2,\ldots,n_L}=N\Phi_{n_1,n_2,\ldots,n_L},$$
where $N=\sum_{l=1}^Ln_l$. The eigenvalues of $\hat n_l$ are simply zero and
one, and consequently $N$ can take any non negative integer value smaller or equal to $L$. Moreover, using the  CAR, we deduce that
$$\hat
n_l\left(b_l\Phi_{n_1,n_2,\ldots,n_L}\right)=\left\{\begin{array}{lll}
	(n_l-1)(b_l\Phi_{n_1,n_2,\ldots,n_L}),\qquad n_l=1\\
	0,\hspace{4.4cm} n_l=0,\\
\end{array}\right.
$$
and $$\hat n_l\left(b_l^\dagger\Phi_{n_1,n_2,\ldots,n_L}\right)=\left\{\begin{array}{lll}
	(n_l+1)(b_l^\dagger\Phi_{n_1,n_2,\ldots,n_L}),\qquad n_l=0\\
	0,\hspace{4.4cm} n_l=1,\\
\end{array}\right.
$$ for all $l$. The
interpretation  is that $b_l$ and $b_l^\dagger$ are 
\emph{annihilation} and \emph{creation} operators\footnote{In some sense,
	$b_l^\dagger$ is {\bf also} an annihilation operator since, acting on a state with $n_l=1$, it destroys that state: we are trying to
	put together two identical fermions, and this operation is forbidden by the Pauli exclusion principle.}.

Of course, $\Hil_F$ has a finite dimension. In particular, for just one mode of fermions, $dim(\Hil_F)=2$, while $dim(\Hil_F)=4$ if $L=2$. This also implies that, contrarily
to what happens for bosons, see below, all the fermionic operators are bounded and can be represented by finite-dimensional matrices.

The vector $\Phi_{n_1,n_2,\ldots,n_L}$ in (\ref{a4}) defines a \emph{vector (or number) state }
over the algebra $\A$
of the operators over $\Hil_F$ as \be
\omega_{n_1,n_2,\ldots,n_L}(X)= \langle\Phi_{n_1,n_2,\ldots,n_L},X\Phi_{n_1,n_2,\ldots,n_L}\rangle, \label{a5} \en where $\langle\,,\,\rangle$
is the scalar product in  $\Hil_F$, and $X\in\A$. These states are used to \emph{project} from quantum to classical dynamics and to fix the
\index{Initial conditions}initial conditions of the considered system, in a way which is widely explored in \cite{bagbook,bagbook2,FFF}.

\subsection{Bosonic operators}

An alternative family of relevant operators arises from the {\em canonical commutation relations} (CCR): we say that a set
of operators $\{a_l,\,a_l^\dagger, l=1,2,\ldots,L\}$, acting on the Hilbert space $\Hil$, satisfy the CCR, if the following hold:\be
[a_l,a_n^\dagger]=\delta_{ln}\Id,\hspace{8mm} [a_l,a_n]=[a_l^\dagger,a_n^\dagger]=0, \label{bno3}\en for all $l,n=1,2,\ldots,L$, $\Id$ being
the identity operator on $\Hil$, and $[x,y]:=xy-yx$.  These operators, for which we refer to \cite{mer,mess} for
instance, are those which are used to describe $L$ different {\em modes} of bosons. As before we can
construct self-adjoint operators $\hat n_l=a_l^\dagger a_l$ and $\hat N=\sum_{l=1}^L \hat n_l$ and, as for fermions, $\hat n_l$ is the
{\em number operator } for the $l$-th mode, while $\hat N$ is the {\em global number operator}.

An o.n. \index{Basis!orthonormal}basis of $\Hil$ can be constructed as follows: we introduce the \index{Vacuum}{\em vacuum} of
the theory, that is a vector $\varphi_{\bf 0}$ which is annihilated by all the operators $a_l$: $a_l\varphi_{\bf 0}=0$ for all
$l=1,2,\ldots,L$. Then we act on $\varphi_{\bf 0}$ with the  operators $a_l^\dagger$ and with their powers, \be
\varphi_{n_1,n_2,\ldots,n_L}:=\frac{1}{\sqrt{n_1!\,n_2!\ldots n_L!}}(a_1^\dagger)^{n_1}(a_2^\dagger)^{n_2}\cdots
(a_L^\dagger)^{n_L}\varphi_{\bf 0}, \label{bno4}\en $n_l=0,1,2,\ldots$, for all $l$, and we normalize the vectors obtained in this way. The set
of the $\varphi_{n_1,n_2,\ldots,n_L}$'s in (\ref{bno4}) forms a complete and o.n. set in $\Hil$, and they are \index{Eigenstates}eigenstates of both $\hat n_l$ and $\hat N$:
$$\hat
n_l\varphi_{n_1,n_2,\ldots,n_L}=n_l\varphi_{n_1,n_2,\ldots,n_L}$$ and
$$\hat N\varphi_{n_1,n_2,\ldots,n_L}=N\varphi_{n_1,n_2,\ldots,n_L},$$
where $N=\sum_{l=1}^Ln_l$. Hence, $n_l$ and $N$ are eigenvalues of $\hat n_l$ and $\hat N$ respectively, none of which necessarily bounded from above. Moreover, using the  CCR we deduce
that
$$\hat
n_l\left(a_l\varphi_{n_1,n_2,\ldots,n_L}\right)=(n_l-1)(a_l\varphi_{n_1,n_2,\ldots,n_L}),$$ for $n_l\geq1$ while, if $n_l=0$, $a_l$ annihilates
the vector, and $$\hat
n_l\left(a_l^\dagger\varphi_{n_1,n_2,\ldots,n_L}\right)=(n_l+1)(a_l^\dagger\varphi_{n_1,n_2,\ldots,n_L}),$$ for all $l$ and for all $n_l$. For
these reasons, if the $L$ different modes of bosons of $\Sc$ are described by $\varphi_{n_1,n_2,\ldots,n_L}$, this means that $n_1$ bosons are in the first mode, $n_2$ in the second mode, and so
on. The operator $\hat n_l$ acts on
$\varphi_{n_1,n_2,\ldots,n_L}$ and returns $n_l$, which is exactly the number of bosons in the l-th mode. The operator $\hat N$ counts the
total number of bosons. Moreover, the operator $a_l$ destroys a boson in the l-th mode, if there is at least one. Otherwise $a_l$ simply
destroys the state. Its adjoint, $a_l^\dagger$, creates a boson in the same mode.
This is why $a_l$ and $a_l^\dagger$ are usually called the {\em annihilation} (or \emph{lowering}) and the
{\em creation} (or {\em raising}) operators.

\vspace{2mm}

The vector $\varphi_{n_1,n_2,\ldots,n_L}$ in (\ref{bno4}) defines a \index{State!vector}{\em vector (or number) state } over the set $\A$ of the operators on $\Hil$ as
\be\omega_{n_1,n_2,\ldots,n_L}(X)= \langle\varphi_{n_1,n_2,\ldots,n_L},X\varphi_{n_1,n_2,\ldots,n_L}\rangle,\label{bno5}\en where
$\langle\,,\,\rangle$ is the scalar product in $\Hil$, and $X\in\A$. As for fermions, these states are used to { project} from quantum to classical
dynamics.

We end this section by recalling that the CAR and CCR  have been extensively used during the past two decades  in the construction of several models in Finance, Decision Making, Biology, Ecology, Sociology,..., and in the analysis of their dynamical behaviour. We refer to \cite{bagbook,bagbook2,FFF}, where many of these examples are discussed in details, and where more references can be found. In particular, we want to mention applications to Decision Making in various contexts, from love affairs, \cite{bagoli,bag5}, to politics, \cite{Bagarello2015b,baggarg}, to game theory and to the analysis of the role of information, \cite{baghavkhr,baggame}.

\section{New ladder operators}\label{sect3}

In this section we introduce a different family of ladder operators obeying certain closed commutation rules which have not been considered so far in the context of macroscopic systems which are interesting for us, and which turn out to simplify a lot our {\em quantum-like} approach, opening the way to new applications.

We start introducing the one-dimensional, self-adjoint, position and momentum operators $\hat x$ and $\hat p=-i\partial_x$, acting on $\Hil=\ltwo$, the Hilbert space of the square integrable functions. They satisfy the commutation rule $[\hat x,\hat p]=i\1$. We then introduce the unitary operator $T=e^{i\hat p}$. The following (equivalent) commutators are satisfied
\be
[T,\hat x]=T, \qquad [T^\dagger,\hat x]=-T^\dagger.
\label{31}\en
Moreover, if $f(x)\in\ltwo$, $Tf(x)=f(x+1)$. Hence $T$ is a translation operator, like its adjoint $T^\dagger$: $T^\dagger f(x)=f(x-1)$. This implies that $T^\dagger=T^{-1}$: $T$ is unitary. 

\vspace{2mm}

{\bf Remark:--} It is clear that $T$ and $T^{-1}$ are not really {\em new} operators, since the role of translation operators in quantum mechanics, many-body, relativity (and many other fields of Physics) is very well known. What is new, we believe, is their role in the analysis of certain macroscopic systems as those which are relevant in this paper, and their nature of ladder-like operators.

\vspace{2mm}

Let us now take a function $\varphi_0(x)\in\ltwo$ such that
$$
<\hat x>_{\varphi_0}=\pin{\varphi_0}{x\varphi_0}=\int_{\mathbb{R}}x|\varphi_0(x)|^2\,dx=0.
$$
Of course any even square-integrable function has this property. One which is particularly easy to use, for our purposes, is the gaussian $\varphi_0(x)=\frac{1}{\pi^{1/4}}e^{-x^2/2}$, which is normalized to one: $\|\varphi_0\|^2=\pin{\varphi_0}{\varphi_0}=1$. From now on we will work with this function, keeping in mind that other choices are also possible. In particular, different choices of $\varphi_0(x)$ could be useful in the attempt to fit experimental data. Let us call 
\be
\varphi_k(x)={T^\dagger}^k\varphi_0(x)=\varphi_0(x-k),
\label{32}\en
$k=0,1,2,3,\ldots$. We get
$$
<\hat x>_{\varphi_k}=\int_{\mathbb{R}}x|\varphi_k(x)|^2\,dx=k.
$$
It is clear that these results can be easily extended to higher dimensions. For instance, in two dimensions we introduce $\hat x_j$, $\hat p=-i\partial_j$, $T_j=e^{i\hat p_j}$, $j=1,2$, and $\varphi_{0,0}(x_1,x_2)=\frac{1}{\sqrt{\pi}}e^{-(x_1^2+x_2^2)/2}=\varphi_0(x_1)\varphi_0(x_2)$, and we have, in particular,
\be
[T_j,\hat x_k]=T_j\delta_{j,k}, \qquad {T_1^\dagger}^{k_1}{T_2^\dagger}^{k_2}\varphi_{0,0}(x_1,x_2)=\varphi_{0,0}(x_1-k_1,x_2-k_2)=:\varphi_{k_1,k_2}(x_1,x_2),
\label{33}\en
as well as
\be
<\hat x_1>_{\varphi_{k_1,k_2}}=\int_{\mathbb{R}^2}x_1|\varphi_{k_1,k_2}(x_1,x_2)|^2\,dx_1\,dx_2=k_1,
\label{34}\en
and
\be
<\hat x_2>_{\varphi_{k_1,k_2}}=\int_{\mathbb{R}^2}x_2|\varphi_{k_1,k_2}(x_1,x_2)|^2\,dx_1\,dx_2=k_2.
\label{35}\en
Then, the comparison with what described in Section \ref{sect2} is easy: $k_j$ are the analogous of the occupation numbers we met before, i.e. the eigenvalues of the (fermionic or bosonic) number operators $\hat n_j$. For this reason we restrict here to $k_j\geq0$ in (\ref{33}), even if this is not really required from a mathematical point of view. Hence, since we go down and up using $T_j$ and $T_j^\dagger$, see (\ref{33}), these latter are like  lowering and a raising operators, respectively (with no vacuum, in principle). The role of $\hat n_j$ is now played by $\hat x_j$, and the vectors in (\ref{a4}) and (\ref{bno4}) are replaced by $\varphi_{k_1,k_2}(x_1,x_2)$, here. We stress that these functions are not mutually orthogonal, while those in Section \ref{sect2} are. Following the general approach discussed in \cite{bagbook}, and later in \cite{bagbook2,FFF}, we need to compute the mean values of the time evolution of the operators $\hat x_j$, on a vector which represents, as usual, the state of the physical system at $t=0$, $\Psi_0$: $\pin{\psi_0}{\hat x_j(t)\Psi_0}$. Of course, the computation of these mean values imply that we have to compute first $\hat x_j(t)$. We will show how this can be done in the next sections, for some explicit applications.

We conclude this section by stressing that the operators $T_j$, despite being some sort of ladder operators, do not obey the CCR or the CAR. Indeed we have
\be
[T_j,T_k^\dagger]=0, \qquad \{T_j,T_k^\dagger\}=2\1\delta_{j,k}, \quad \mbox{ with } \quad  T_j^2\neq 0.
\label{36}\en
In other words, we are not dealing, here, with bosons or with fermions. Still, because of (\ref{33}), we have a {\em ladder structure} which can be used to analyse some concrete systems were ladder operators have already been adopted with success. As we will see later, this different choice of operators will help us to simplify the technical difficulties we have mentioned before, arising when using CAR or CCR. 

\section{A quadratic predator-prey model}\label{sect4}

In \cite{bagbook} we have considered a simple predator-prey model in terms of fermionic operators. In particular, we have considered the following self--adjoint Hamiltonian $H$ for the system $\Sc$: \be
H=H_0+\lambda H_I,\qquad H_0=\omega_1a_1^\dagger a_1+\omega_3a_3^\dagger a_3, \quad H_I=a_1^\dagger a_3+a_3^\dagger a_1. \label{41}\en Here
$\omega_j$ and $\lambda$ are real positive quantities, to ensure that $H$ is self-adjoint.  Of course, if we take $\lambda=0$, then the two populations (1 are the predator, $\tau_1$ and 3 the prey\footnote{This might appear a strange choice, but its rationale will be clarified later on.}, $\tau_3$) do not interact. If $\lambda\neq0$, which is the relevant case for us, $H$ contains also the contribution $H_I=a_1^\dagger a_3+a_3^\dagger a_1$,
whose meaning is the following: $a_1^\dagger a_3$ makes the density of $\tau_1$ to increase (because
of $a_1^\dagger$) and that of $\tau_3$ to decrease (because of $a_3$). The adjoint contribution, $a_3^\dagger a_1$, is responsible for the
opposite phenomenon. The equations of motion are obtained by using the Heisenberg's recipe, $\dot X(t)=i[H,X(t)]$: \be
\begin{aligned}
	&\dot a_1(t)=-i\omega_1 a_1(t)-i\lambda a_3(t),\\
	&\dot a_3(t)=-i\omega_3 a_3(t)-i\lambda a_1(t),
\end{aligned}
\label{42} \en which can be solved with the initial conditions $a_1(0)=a_1$ and $a_3(0)=a_3$. We refer to \cite{bagbook} for the details of the solution, and for their use in computing the time dependence of the {\em density operators\footnote{These should not be confused with density matrices which, in a purely quantum mechanical context, have very specific properties.}}  $n_j(t):=\left<\varphi_{n_1,n_3},
\hat n_j(t)\varphi_{n_1,n_3}\right>$ of the two species, where $\hat n_j(t)=a_j^\dagger(t)a_j(t)$. What is relevant here is that, defining
\[
\begin{aligned}
	&\delta=\sqrt{(\omega_1-\omega_3)^2+4\lambda^2},\\
	&\Phi_+(t)=2\exp\left(-\frac{it(\omega_1+\omega_3)}{2}\right)\cos\left(\frac{\delta t}{2}\right),\\
	&\Phi_-(t)=-2i\exp\left(-\frac{it(\omega_1+\omega_3)}{2}\right)\sin\left(\frac{\delta t}{2}\right),
\end{aligned}
\]
and assuming that $\delta\neq0$, we obtain \be
n_1(t)=n_1\frac{(\omega_1-\omega_3)^2}{(\omega_1-\omega_3)^2+4\lambda^2}+ \frac{4\lambda^2}{(\omega_1-\omega_3)^2+4\lambda^2}
\left\{n_1\cos^2\left(\frac{\delta t}{2}\right)+n_3\sin^2\left(\frac{\delta t}{2}\right)\right\}, \label{43}\en and \be
n_3(t)=n_3\frac{(\omega_1-\omega_3)^2}{(\omega_1-\omega_3)^2+4\lambda^2}+ \frac{4\lambda^2}{(\omega_1-\omega_3)^2+4\lambda^2}
\left\{n_3\cos^2\left(\frac{\delta t}{2}\right)+n_1\sin^2\left(\frac{\delta t}{2}\right)\right\}. \label{44}\en Notice that these formulas
automatically imply that $n_1(t)+n_3(t)=n_1+n_3$, independently of $t$ and $\lambda$. This is expected, since it is easy to check that $[H,\hat
n_1+\hat n_3]=0$. Hence, the total density of the two species is preserved during the time evolution, even in presence of interaction (i.e.,
when $\lambda\neq0$). What is relevant for us here is that we get an oscillatory behaviour of the two species $\tau_1$ and $\tau_3$, and that these oscillations are in opposition of phase: when $n_1(t)$ increases, $n_2(t)$ decreases, and vice-versa. This is what we expect from a dynamical system like ours, and was considered in \cite{bagoli} as a good indication of the relevance of quantum dynamics also in a classical context. 

\vspace{2mm}

{\bf Remark:--} It is useful to stress that the same Hamiltonian as in (\ref{41}) was used in a different contexts. In particular, it was used in a population dynamics model, \cite{BO2013,BGO2015}, and in the analysis of a love affair, \cite{bagoli}. In this latter case, however, the fermionic operators were replaced by bosonic ones.

\subsection{What now?}

What is interesting for us here, first of all,  is to show that it is possible to recover the same qualitative results as above adopting the new ladder operators $\hat x_j$ and $T_j$. 

The counterpart of the Hamiltonian (\ref{41}) is easily written. It is sufficient to replace the fermionic number operators $\hat n_j=a_j^\dagger a_j$ with $\hat x_j$, and the {\em old} ladder $a_j$ with the {\em new} ones, $T_j$. Hence we get:
\be
\tilde H=\omega_1\hat x_1+\omega_3\hat x_3+\lambda\left(T_1^\dagger T_3+T_3^\dagger T_1\right).
\label{45}\en
Here, as in (\ref{41}), $1$ is the label of predators and 3 that of the prey. Because of the (\ref{36}), the Heisenberg equations of motion for the system are
\be
\left\{
\begin{array}{ll}
	\frac{d}{dt}\hat x_1(t)=i\lambda\left(T_1(t)T_3^\dagger(t)-T_3(t)T_1^\dagger(t)\right) \\
	\frac{d}{dt}\hat x_3(t)=-i\lambda\left(T_1(t)T_3^\dagger(t)-T_3(t)T_1^\dagger(t)\right) \\
	\frac{d}{dt}T_1(t)=-i\omega_1 T_1(t), \\
	\frac{d}{dt}T_3(t)=-i\omega_3 T_3(t). \\
\end{array}
\right. \label{46} \en
It is clear then that $\hat x_1(t)+\hat x_3(t)$ is a constant of motion. This is in agreement with what we have found in the fermionic model, as commented after (\ref{44}). We further observe that the equations for $T_j(t)$ in (\ref{46}) can be solved easily. Let us assume, for the moment, that $\omega_j\neq0$, and that $\Omega=\omega_1-\omega_3\neq0$, too. In this situation the differential equations for $\hat x_j(t)$ produce the following solutions:
\be
\left\{
\begin{array}{ll}
	\hat x_1(t)=\hat x_1(0)-\frac{\lambda}{\Omega}\left(\left(e^{-i\Omega t}-1\right)T_1T_3^\dagger-\left(e^{i\Omega t}-1\right)T_3T_1^\dagger\right) \\
	\hat x_3(t)=\hat x_3(0)+\frac{\lambda}{\Omega}\left(\left(e^{-i\Omega t}-1\right)T_1T_3^\dagger-\left(e^{i\Omega t}-1\right)T_3T_1^\dagger\right). \\
\end{array}
\right. \label{47} \en
If we next compute their mean values on a vector $\varphi_{k_1,k_3}(x_1,x_3)$, $x_j(t)$, $j=1,3$, the result is the following:
\be
\left\{
\begin{array}{ll}
	x_1(t)=k_1-\frac{2\lambda}{\Omega\sqrt{e}}\left(\cos(\Omega t)-1\right) \\
	x_3(t)=k_3+\frac{2\lambda}{\Omega\sqrt{e}}\left(\cos(\Omega t)-1\right), \\
\end{array}
\right. \label{48} \en
where we have used also $\pin{\varphi_{1,0}}{\varphi_{0,1}}=\frac{1}{\sqrt{e}}$. This is what we expected, and it is also in agreement with the fermionic model: the two populations oscillate, around their initial values, in opposition of phases. This result suggests that these new ladder operators can be indeed relevant in the analysis of macroscopic systems, and can be a valid alternative to bosonic and fermionic creation and annihilation operators.

\vspace{2mm}

{\bf Remarks:--} (1) It is interesting to see what happens if $\omega_1=\omega_3$. In this case $\Omega=0$ and $\hat x_j(t)$, rather than being oscillating in time, become constant and, of course, $\hat x_1(t)+\hat x_3(t)$ is preserved a fortiori. This implies, in particular, that the same oscillations as for the fermionic models are recovered only in presence of the {\em free term} $\omega_1\hat x_1+\omega_3\hat x_3$ in $\tilde H$, and if $\omega_1$ and $\omega_3$ are different. This is indeed different from what we observe in (\ref{43})-(\ref{44}): in the present settings, in fact, the presence of a free Hamiltonian, with different {\em frequencies} for the different agents, is the key ingredient of the model to have an oscillatory behavior. 

(2) It may be useful to notice here that, without any further assumption, $x_1(t)$ and $x_3(t)$ in (\ref{48}) are not positive definite. However, it is not difficult to find conditions on $k_1$, $k_3$, $\lambda$ and $\Omega$ which ensure that $x_j(t)\geq0$, $j=1,3$, $\forall t$. This is important in our particular system, since $x_j(t)$ describes the density of the species $j$, which cannot be negative. However, for different systems, different interpretations can also be relevant and therefore the fact that the operators $\hat x_j(t)$ are not necessarily positive, rather than being a problem, could be seen as a interesting feature of the present approach, when compared with what is deduced using CCR or CAR.

\section{A cubic predator-prey model}\label{sect5}

What we will show in this section is how the above model can be extended by introducing a third agent ($\tau_2$) and its operators ($\hat x_2$, $T_2$ and $T_2^\dagger$), producing again an exactly solvable (non linear) set of differential equations, despite of the fact that the Hamiltonian below is cubic. 

We introduce here
\be
H=\omega_1\hat x_1+\omega_2\hat x_2+\omega_3\hat x_3+\lambda_1\left(T_1^\dagger T_2T_3+T_1T_2^\dagger T_3^\dagger\right)+\lambda_2\left(T_1^\dagger T_2^\dagger T_3+T_1T_2 T_3^\dagger\right).
\label{51}\en
We see that $H$ has a free term, $H_0=\omega_1\hat x_1+\omega_2\hat x_2+\omega_3\hat x_3$, and an interaction, $H_1=\lambda_1\left(T_1^\dagger T_2T_3+T_1T_2^\dagger T_3^\dagger\right)+\lambda_2\left(T_1^\dagger T_2^\dagger T_3+T_1T_2 T_3^\dagger\right)= H_{1,a}+H_{1,b}$, which is the sum of two contributions, whose meaning we are going to discuss soon. First of all, to fix the ideas, we observe that, as before, {\em mode 1} is attached to predators, while {\em mode 3} is the prey' mode. Mode 2 is related to a third agent which we introduce to mediate the interaction between predators and prey. In particular, from the analytical expression of $H_1$, we observe that both in $H_{1,a}$ and in $H_{1,b}$, we have $T_1^\dagger T_3$ or $T_1T_3^\dagger$. This describes the fact that when the predators increase, the prey decrease, and vice-versa. The difference between $H_{1,a}$ and $H_{1,b}$ is in the role of $T_2$, and in the difference between $\lambda_1$ and $\lambda_2$. For instance, while in $H_{1,a}$ we have $T_1^\dagger T_2T_3$, in $H_{1,b}$ the term $T_1^\dagger T_2^\dagger T_3$ appears. In fact, we could rewrite
\be
H_1=T_1^\dagger (\lambda_1T_2+\lambda_2 T_2^\dagger)T_3+T_1 (\lambda_1T_2^\dagger+\lambda_2 T_2)T_3^\dagger,
\label{52}\en
which clarifies the difference of this model with respect to the one in (\ref{45}): what we are considering here is still a predator-prey model as the one in Section \ref{sect4}, but with coefficients which depend on a third agent and, as such, introduce an extra effective time dependence in the game, as it will be clear soon: so the interaction between predators and prey is {\em effectively time dependent}.

The initial state of the system is, extending what discussed before, $\varphi_{k_1,k_2,k_3}(x_1,x_2,x_3)=\varphi_{0,0,0}(x_1-k_1,x_2-k_2,x_3-k_3)$, where
$$
\varphi_{0,0,0}(x_1,x_2,x_3)=\frac{1}{\pi^{3/2}}e^{-(x_1^2+x_2^2+x_3^2)/2}=\varphi_0(x_1)\varphi_0(x_2)\varphi_0(x_3).
$$
As always, $k_1$, $k_2$ and $k_3$ describe the initial status of the densities of the three agents, in analogy with (\ref{34}) and (\ref{35}).

The Heisenberg equations of motion for the system are
\be
\left\{
\begin{array}{ll}
	\frac{d}{dt}\hat x_1(t)=i\lambda_1\left(T_1(t)T_2^\dagger(t)T_3^\dagger(t)-T_1^\dagger(t)T_2(t)T_3(t)\right)+i\lambda_2\left(T_1(t)T_2(t)T_3^\dagger(t)-T_1^\dagger(t)T_2^\dagger(t)T_3(t)\right) \\
	\frac{d}{dt}\hat x_2(t)=i\lambda_1\left(T_1^\dagger(t)T_2(t)T_3(t)-T_1(t)T_2^\dagger(t)T_3^\dagger(t)\right)+i\lambda_2\left(T_1(t)T_2(t)T_3^\dagger(t)-T_1^\dagger(t)T_2^\dagger(t)T_3(t)\right)  \\
	\frac{d}{dt}\hat x_3(t)=i\lambda_1\left(T_1^\dagger(t)T_2(t)T_3(t)-T_1(t)T_2^\dagger(t)T_3^\dagger(t)\right)+i\lambda_2\left(T_1^\dagger(t)T_2^\dagger(t)T_3(t)-T_1(t)T_2(t)T_3^\dagger(t)\right)  \\
	\frac{d}{dt}T_j(t)=-i\omega_j T_j(t), \qquad j=1,2,3. \\
\end{array}
\right. \label{53} \en
Due to the equations for $T_j(t)$, we can easily solve also those for $\hat x_j(t)$, and we get, after some easy calculations,
$$
\hat x_1(t)=\hat x_1(0)-\frac{\lambda_1}{\Omega_1}\left(T_1T_2^\dagger T_3^\dagger (e^{-i\Omega_1t}-1)+T_1^\dagger T_2T_3(e^{i\Omega_1t}-1)\right)+
$$
\be
-\frac{\lambda_2}{\Omega_2}\left(T_1T_2 T_3^\dagger (e^{-i\Omega_2t}-1)+T_1^\dagger T_2^\dagger T_3(e^{i\Omega_2t}-1)\right),
\label{54}\en
$$
\hat x_2(t)=\hat x_2(0)+\frac{\lambda_1}{\Omega_1}\left(T_1^\dagger T_2 T_3 (e^{i\Omega_1t}-1)+T_1 T_2^\dagger T_3^\dagger(e^{-i\Omega_1t}-1)\right)+
$$
\be
-\frac{\lambda_2}{\Omega_2}\left(T_1T_2 T_3^\dagger (e^{-i\Omega_2t}-1)+T_1^\dagger T_2^\dagger T_3(e^{i\Omega_2t}-1)\right),
\label{55}\en
and
$$
\hat x_3(t)=\hat x_3(0)+\frac{\lambda_1}{\Omega_1}\left(T_1^\dagger T_2 T_3 (e^{i\Omega_1t}-1)+T_1 T_2^\dagger T_3^\dagger(e^{-i\Omega_1t}-1)\right)+
$$
\be
+\frac{\lambda_2}{\Omega_2}\left(T_1^\dagger T_2^\dagger T_3 (e^{i\Omega_2t}-1)+T_1 T_2 T_3^\dagger(e^{-i\Omega_2t}-1)\right).
\label{56}\en
Here we have introduced 
\be
\Omega_1=\omega_1-\omega_2-\omega_3, \qquad \Omega_2=\omega_1+\omega_2-\omega_3,
\label{57}\en
and the solution is found under the assumption that $\Omega_1$ and $\Omega_2$ are both not zero. As we have commented before, the solution could also be easily deduced even if this constraint is not satisfied, but looks different from the one above. This case will not be considered here.

From formulas (\ref{54})-(\ref{56}) it is easy to see that $\hat x_1(t)+\hat x_3(t)$ is a constant of motion, as in the simple two-components model in Section \ref{sect4}. This is reasonable, since the Hamiltonian in (\ref{51}) again describes the fact that, when the predators increase, prey decreases, and vice-versa. This is independent of the second agents, and in fact $\hat x_2(t)$ does not enter in the expression of the integral of motion. We could reach the same conclusion also simply by looking for values of $\alpha_j$ such that $I=\sum_{j=1}^3\alpha_j\hat x_j$ commutes with $H$. The computation is easy and the only solution is that $[H,I]=0$ if $\alpha_2=0$ and $\alpha_1=\alpha_3$. Of course, if we take $\alpha_1=\alpha_3=1$, we go back to what we have just deduced.

The last step consists in computing the mean values of the operators $\hat x_j(t)$ on  $\Psi(0)=\varphi_{k_1,k_2,k_3}(x_1,x_2,x_3)$, the initial state of the system. After some computations we get the following:
\be
\left\{
\begin{array}{ll}
	x_1(t)=k_1-2e^{-3/4}\left(\frac{\lambda_1}{\Omega_1}(\cos(\Omega_1t)-1)+\frac{\lambda_2}{\Omega_2}(\cos(\Omega_2t)-1)\right)\\
	x_2(t)=k_2+2e^{-3/4}\left(\frac{\lambda_1}{\Omega_1}(\cos(\Omega_1t)-1)-\frac{\lambda_2}{\Omega_2}(\cos(\Omega_2t)-1)\right)\\
	x_3(t)=k_3+2e^{-3/4}\left(\frac{\lambda_1}{\Omega_1}(\cos(\Omega_1t)-1)+\frac{\lambda_2}{\Omega_2}(\cos(\Omega_2t)-1)\right).
\end{array}
\right. 
\label{58}\en
We introduce now the function
\be
V(t)=\frac{\lambda_1}{\Omega_1}(\cos(\Omega_1t)-1)+\frac{\lambda_2}{\Omega_2}(\cos(\Omega_2t)-1),
\label{59}\en
which contains all the oscillations of both $x_1(t)$ and $x_3(t)$. Of course, the smaller these oscillations, the closer the system is to a sort of equilibrium between prey and predators. This suggests that the interaction between the species is small. On the other hand, if we have larger oscillations, the two species strongly interact. Of course, our analysis depend on the sign of $\Omega_1$ and $\Omega_2$. Notice that, because of (\ref{57}), and recalling that each $\omega_j$ is strictly positive, then we have $\Omega_2>\Omega_1$. This implies that the only possibilities are the following: (i) $\Omega_1>0$ (and $\Omega_2>0$ automatically); (ii) $\Omega_1<0$ and $\Omega_2<0$; (iii) $\Omega_1<0$ and $\Omega_2>0$. It is a simple exercise to check that the amplitude of the oscillations in all these cases is the following:
\be
\Delta=\max_t(V(t))-\min_t(V(t))=2\left(\frac{\lambda_1}{|\Omega_1|}+\frac{\lambda_2}{|\Omega_2|}\right).
\label{510}\en 
This formula shows that $x_1(t)$ and $x_3(t)$ present large oscillations if at least one between $\frac{\lambda_1}{|\Omega_1|}$ and $\frac{\lambda_2}{|\Omega_2|}$ is big. On the other hand, if we want to have some kind of equilibrium between the predators and the prey, i.e. if we want the oscillations of these two species to stay small, we need to have both the above ratios small. This is possible for small values of the $\lambda_j$'s and/or for large values of $|\Omega_j|$'s. In particular, the larger $|\omega_1-\omega_3|$ is, with respect to $\omega_2$, the smaller the oscillations (for $\lambda_j$ fixed).

\begin{figure}[th]
	\begin{center}
		\includegraphics[width=0.47\textwidth]{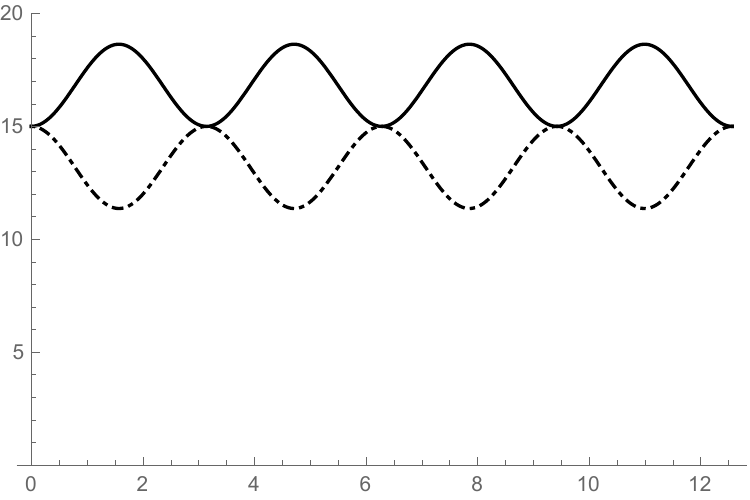}
		\hfill\\[20pt] \includegraphics[width=0.47\textwidth]{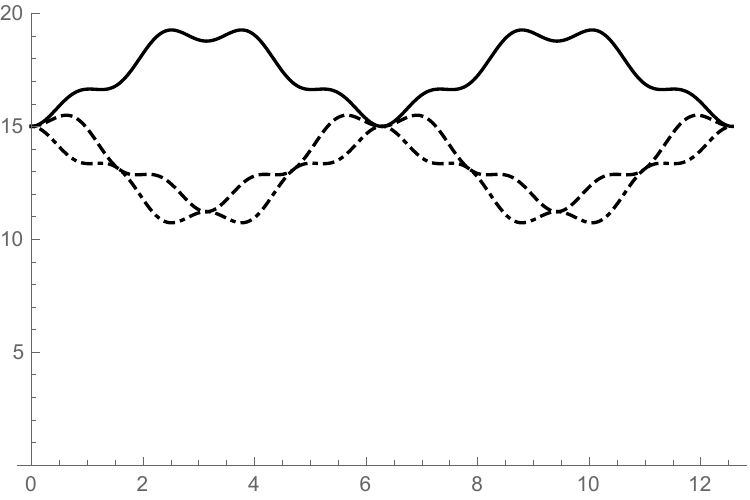}%
		\hspace{7mm}\includegraphics[width=0.47\textwidth]{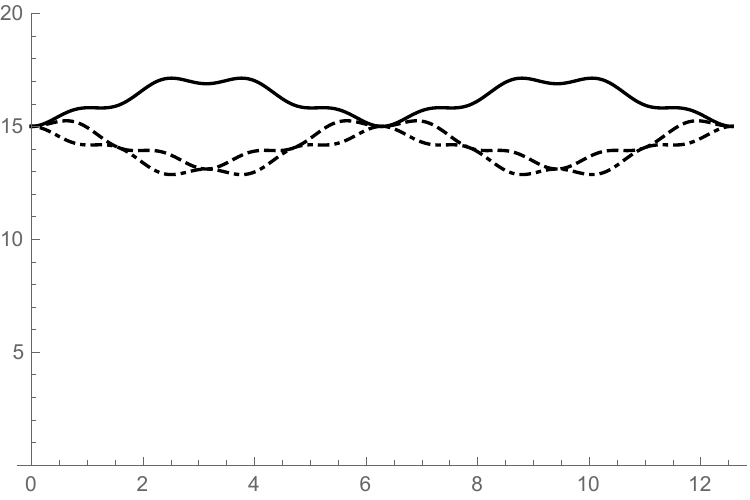}
	\end{center}
	\caption{{\protect\footnotesize $x_1(t)$ (continuous line) and $x_3(t)$ (dot-dashed line) for the two agents model (up) and $x_1(t)$ (continuous line), $x_2(t)$ (dashed line) and $x_3(t)$ (dot-dashed line) for the three agents model (down). Constants are as follows: $\lambda=3$, $\Omega=2$ (up); $\lambda_1=2$, $\lambda_2$, $\Omega_1=2$, $\Omega_2=3$ (down, left); $\lambda_1=2$, $\lambda_2$, $\Omega_1=1$, $\Omega_2=4$ (down, right).}}
	\label{fig1}
\end{figure}

From Figure \ref{fig1} we see, first of all, the difference between the two Hamiltonians considered so far, in (\ref{45}) and (\ref{51}): in the first line we have {\em simple} oscillations, out of phase, which describe a simple oscillatory behavior of the two species. In the bottom line the time evolution of the two densities of predators and prey again oscillate out of phase, but in a more {\em elaborate} way. The reason for that is in the presence of the third species ($x_2(t)$ in the figure), whose values are always in the range $[x_3(t),x_1(t)]$, and stay here much closer to $x_3(t)$ than to $x_1(t)$. Going back to (\ref{52}), Figure \ref{fig1} suggests to interpret, as already observed, $\tau_2$ as a sort of time-dependent environment which changes with time the strength of the interaction between $\tau_1$ and $\tau_3$, and this {\em effective time dependence} is reflected in the different behavior between the plots in the up and low pictures in Figure \ref{fig1}. 

\vspace{2mm}

We have seen here that the Hamiltonians considered in (\ref{45}) and (\ref{51}) give rise to exactly solvable models. This is what happens also if we consider a $N$-agents system, $N\geq2$, at least under the assumption that its dynamics is driven by an Hamiltonian
$
h=h_0+h_I
$, where $h_0=\sum_{j=1}^N\omega_j\hat x_j$ and $h_I=\sum_{\bf k}\alpha_{\bf k}T_1^\sharp\cdots T_{k_N}^\sharp$, where $T_j^\sharp$ is either $T_j$ or $T_j^\dagger$, and $\alpha_{\bf k}$ are coefficients such that $h_I=h_I^\dagger$, some of which could be zero. With this choice, we deduce a set of differential equations which extends that in (\ref{53}), and the solution we get will simply be a slightly longer version of (\ref{58}): again, a periodic or a quasi-periodic dynamics.

\section{Another cubic predator-prey model }\label{sect6}

In view of what we have seen so far, it might be interesting to show that the operators $(\hat x_j, T_k, T^\dagger_k)$ can produce, when used properly in the definition of certain Hamiltonians, some dynamics which is not necessarily periodic or quasi-periodic. This is what we will
discuss in this section, proposing  a different cubic version of the predator-prey considered in Section \ref{sect4} in which, rather than considering a third agent, we modify the interaction Hamiltonian in (\ref{45}) in order to take into account the possibility that, if there is no predator in the system, there is no interaction with the prey, as it is natural to expect.  This can be described by the Hamiltonian 
\be
H=\omega_1\hat x_1+\omega_3\hat x_3+\lambda\left(\hat x_1T_1^\dagger T_3+T_1T_3^\dagger \hat x_1\right),
\label{61}\en
where, as always, $\omega_1$ and $\omega_3$ are (strictly) positive, and $\lambda$ is real. In fact, the presence of $\hat x_1$ in the interaction term, $\hat x_1T_1^\dagger T_3+T_1T_3^\dagger \hat x_1$, guarantees that this interaction does not contribute in absence of predators. Of course, the same could be achieved replacing $\hat x_1T_1^\dagger T_3+T_1T_3^\dagger \hat x_1$ with $\hat x_3T_1^\dagger T_3+T_1T_3^\dagger \hat x_3$. In this case we would have no interaction between predators and prey in absence of prey. Incidentally we observe that, in view of this interpretation, there is no reason to consider an interaction term in which both $\hat x_1$ and $\hat x_3$ appear. Going back to the $H$ in (\ref{61}), we deduce the following set of differential equations:
\be
\left\{
\begin{array}{ll}
	\frac{d}{dt}\hat x_1(t)=i\lambda\left(T_1(t)\hat x_1(t)T_3^\dagger(t)-\hat x_1(t)T_1^\dagger(t)T_3(t)\right) \\
	\frac{d}{dt}\hat x_3(t)=-i\lambda\left(T_1(t)\hat x_1(t)T_3^\dagger(t)-\hat x_1(t)T_1^\dagger(t)T_3(t)\right)  \\
	\frac{d}{dt} T_1(t)=-i\omega_1 T_1(t)-i\lambda T_3(t)-i\lambda T_1^2(t)T_3^\dagger(t)\\
	\frac{d}{dt}T_3(t)=-i\omega_3 T_3(t). \\
\end{array}
\right. \label{62} \en
The first obvious consequence is that $\hat x_1(t)+\hat x_3(t)$ is an integral of motion, exactly as in Section \ref{sect4}. This is not surprising, since the Hamiltonian in (\ref{61}) describes exactly the same effect as that in (\ref{45}), with the only (major) difference that the changes in the densities of predators and prey, here, occur only in presence of predators. Otherwise $H$ simply does not change the status of the system. A second consequence of (\ref{62}) is that the dynamics of $T_1$ and $T_3$ is very different. We have, indeed, $T_3(t)=e^{-i\omega_3t}T_3$, while the equation for $T_1(t)$ is much more complicated. Summarizing, the equations in (\ref{62}) reduce now to
\be
\left\{
\begin{array}{ll}
	\frac{d}{dt}\hat x_1(t)=i\lambda\left(e^{i\omega_3t}T_1(t)\hat x_1(t)T_3^\dagger-e^{-i\omega_3t}\hat x_1(t)T_1^\dagger(t)T_3\right) \\
	\frac{d}{dt} T_1(t)=-i\omega_1 T_1(t)-i\lambda  e^{-i\omega_3t}T_3-i\lambda e^{i\omega_3t} T_1^2(t)T_3^\dagger,\\
\end{array}
\right. \label{63} \en
while $\hat x_3(t)$ can be easily deduced from $\hat x_1(t)$.

\subsection{An interlude: an approximated solution}

This system of equations can be easily solved using a perturbative approach in $\lambda$. At the zero-th order in $\lambda$ system (\ref{63}) reduces to
$$
\left\{
\begin{array}{ll}
	\frac{d}{dt}\hat x_1^{(0)}(t)=0\\
	\frac{d}{dt} T_1^{(0)}(t)=-i\omega_1 T_1^{(0)}(t),\\
\end{array}
\right. 
$$
whose solution is trivial: $\hat x_1^{(0)}(t)=\hat x_1(0)=\hat x_1$ and $T_1^{(0)}(t)=e^{-i\omega_1t}T_1(0)=e^{-i\omega_1t}T_1$. Replacing these solutions in the right-hand side of (\ref{63})  we get the following equation:
$$
\left\{
\begin{array}{ll}
	\frac{d}{dt}\hat x_1^{(1)}(t)=i\lambda\left(e^{-i\Omega t}T_1\hat x_1T_3^\dagger-e^{i\Omega t}\hat x_1T_1^\dagger T_3\right) \\
	\frac{d}{dt} T_1^{(1)}(t)=-i\omega_1e^{-i\omega_1t}T_1-i\lambda  e^{-i\omega_3t}T_3-i\lambda e^{i(\omega_3-2\omega_1)t} T_1^2T_3^\dagger,\\
\end{array}
\right.
$$
where $\Omega=\omega_1-\omega_3$. Solving the first equation under the assumption that $\Omega\neq0$, and computing the mean value of $\hat x_1^{(1)}(t)$ on a vector $\varphi_{k_1,k_2}$ we get, after simple computations
\be
x_1^{(1)}(t)=k_1-\frac{2\lambda}{\Omega\sqrt{e}}\left(k_1+\frac{1}{2}\right)\left(\cos\Omega t-1\right).
\label{64}\en
This solution should be compared with (\ref{48})$_1$. We see that they look very similar. The only difference is in the presence of $k_1+\frac{1}{2}$ in (\ref{64}) which means that the amplitude of the oscillations of $x_1^{(1)}(t)$ depend on the initial conditions on $\hat x_1(t)$, contrarily to what happens in (\ref{48}). This is reasonable (the initial conditions influence the dynamics!), but the result is, in fact, not particularly different from what we get using the model of Section \ref{sect4}: at the first order in $\lambda$ we get, also for the Hamiltonian in (\ref{61}), an oscillatory time evolution of the densities of predators and prey.

\subsection{Looking for an exact solution}

Let us now go back to (\ref{63})$_2$. Introducing the new unknown function $S(t)=e^{i\omega_1 t}T_1(t)$ this equation becomes 
\be
\dot S(t)=-i\lambda e^{i\Omega t} T_2-i\lambda e^{-i\Omega t} T_2^\dagger S^2(t),  
\label{65}\en
which is a sort of operator version of the Riccati equation, see e.g. \cite{arfk}. Using the usual changes of variables one adopts for that class of equations, and using also the fact that here $T_2$ is unitary, we find that
\be
T_1(t)=c_+e^{\beta_+t}+c_-e^{\beta_-t},
\label{66}\en
where
\be
\beta_\pm=\alpha_\pm-i(\omega_1+\Omega), \quad \mbox{ with }\quad \alpha_\pm=\frac{1}{2}\left(-i\Omega\pm\sqrt{4\lambda^2-\Omega^2}\right),
\label{67}\en
and
\be
c_+=\frac{-i}{\sqrt{4\lambda^2-\Omega^2}}\left(\lambda T_2^{-1}+i\alpha_+ T_1\right), \qquad 
c_-=\frac{i}{\sqrt{4\lambda^2-\Omega^2}}\left(\lambda T_2^{-1}+i\alpha_- T_1\right).
\label{68}\en
We observe that these latter are indeed operators. We also observe that the time evolution of $T_1(t)$ is purely oscillatory if $\Re(\alpha_\pm)=0$, i.e. if $4\lambda^2\leq \Omega^2$. In fact, in this situation, we also have if $\Re(\beta_\pm)=0$. On the other hand, if $\Re(\alpha_\pm)\neq0$,  i.e. if $4\lambda^2> \Omega^2$, $\Re(\beta_+)=-\Re(\beta_-)$, and the two exponentials in (\ref{66}) produce an exponential growth and an exponential decay to zero.

If we now insert (\ref{66}) into (\ref{63})$_1$ we get, after several manipulations, 
the equation
\be
\frac{d}{dt}\hat x_1(t)= r(t)+\hat x_1(t)v(t),
\label{69}\en
where the order is important, and where we have introduced 
\be
r(t)=i\lambda\left(d_+e^{(p-iq)t}+d_-e^{-(p+iq)t}\right) .
\label{610}\en
\be
v(t)=i\lambda\left(d_+e^{(p-iq)t}+d_-e^{-(p+iq)t}-d_+^\dagger e^{(p+iq)t}-d_-^\dagger e^{-(p-iq)t}\right).
\label{611}\en
Here we have
\be
p=\frac{1}{2}\sqrt{4\lambda^2-\Omega^2}, \qquad q=\frac{5\Omega}{2}, \qquad d_\pm=c_\pm T_2^\dagger.
\label{612}\en 
Notice that the $d_\pm$ are operators, so that $r(t)$ and $v(t)$ are, in fact, time-dependent operators. Because of the nature of these functions and of $\hat x_1(t)$, it is necessary to pay a particular attention when solving (\ref{69}). However, we can conclude that this lack of abelianity does not cause any particular complication here, and the solution of (\ref{69}) can be found:
\be
\hat x_1(t)=\left(\int r(t)e^{-\int v(t)dt}dt+c\right)e^{\int v(t)dt}dt,
\label{613}\en
where $c$ is a suitable constant of integration, to be fixed by the initial conditions. The mean value of (\ref{613}) on a state defined by $\varphi_{k_1,k_2}$ produces the exact counterpart of (\ref{64}). Not surprisingly, this explicit computation is not so simple, and its details are not so relevant for us and will not be given here. What is more useful to stress is that we are able, using the ladder operators $\hat x_j$ and $T_j$ to construct dynamical systems with a non purely oscillatory dynamics, and (in principle) exactly solvable. This was not so simple for fermionic and bosonic ladder operators, and opens new ways for further applications.

\section{A different view to our systems and conclusions}\label{sect7}

In this paper we have introduced a new set of ladder operators which can be adopted, instead of fermionic or bosonic operators, in the quantum-like description of some macroscopic system. The main difference, with respect to what we have considered in our previous research, is that the computations can be {\em simpler than before} and can be carried out in pure analytical terms, even in presence of non purely quadratic terms in the Hamiltonian, at least of some special kind. 

From the point of view of the applications, what we have considered here are three variations on the same theme: predator-prey models. But, as already commented before, with very little  changes we can adapt the same models to a different situation, i.e. to a love affair between Alice and Bob, in which the rule of the model is the following: the more one agent (e.g., Bob) is attracted by the other (e.g., Alice), the less Alice is attracted by Bob, and vice-versa. The rule is, mutatis mutandis, quite similar to that of the predator-prey system in Section \ref{sect4}. If we further use the model in Section \ref{sect5} in the description of the affair between Alice and Bob, we can interpret the agent $\tau_2$ as a third actor in the affair (families, friends, environments,...) which make Alice and Bob still oscillating, but in a slightly smoother way: friends try to support Alice and Bob's relationship! In this perspective, smaller amplitude in the oscillations in Figure \ref{fig1} correspond to more stable relationships. As for the model in Section \ref{sect6}, this simply introduces a natural constraint: for Alice and Bob to have some interaction, it is necessary to have, first of all, an interest of Alice for Bob (this is the role of $\hat x_1$ in (\ref{61}))! Otherwise, there is no reason to expect any variation in the reciprocal {\em moods} of the two agents.

Summarizing, we can surely claim that these new ladder operators are {\em technically simpler} than those adopted so far by us, and by other authors, and allow to get non trivial dynamical behaviors. Hence, they are surely worthly of a deeper analysis.

\section*{Acknowledgements}

The author acknowledges partial financial support from Palermo University and from G.N.F.M. of the INdAM. This work has also been partially supported by the PRIN grant {\em Transport phenomena in low dimensional
	structures: models, simulations and theoretical aspects}, by Project {\em CAESAR} and by Project {\em ICON-Q}. The author expresses his gratitude to the Referees for their comments.

\end{document}